\documentclass[prl,twocolumn,floatfix,groupedaddress,
nofootinbib,showpacs,preprintnumbers,amsmath,amssymb,amsfonts,superscriptaddress]
{revtex4}
\usepackage{graphicx}
\usepackage{mathrsfs}
\usepackage{bm}

\begin{document}

\title{Garvey-Kelson Relations for Nuclear Charge Radii}
\author{J. Piekarewicz}
\affiliation{Department of Physics, Florida State University,
             Tallahassee, FL {\sl 32306}}
\author{M. Centelles}
\affiliation{Departament d'Estructura i Constituents de la Mat\`eria
and Institut de Ci\`encies del Cosmos, Facultat de F\'{\i}sica,
Universitat de Barcelona, Diagonal {\sl 647}, {\sl E-08028} Barcelona,
Spain} 
\author{X. Roca-Maza}
\affiliation{Departament d'Estructura i Constituents de la Mat\`eria
and Institut de Ci\`encies del Cosmos, Facultat de F\'{\i}sica,
Universitat de Barcelona, Diagonal {\sl 647}, {\sl E-08028} Barcelona,
Spain} 
\author{X. Vi\~nas}
\affiliation{Departament d'Estructura i Constituents de la Mat\`eria
and Institut de Ci\`encies del Cosmos, Facultat de F\'{\i}sica,
Universitat de Barcelona, Diagonal {\sl 647}, {\sl E-08028} Barcelona,
Spain} 

 \date{\today}
 \bigskip
 \begin{abstract}
 \medskip
The Garvey-Kelson relations (GKRs) are algebraic expressions
originally developed to predict nuclear masses. In this letter we 
show that the GKRs provide a fruitful framework for the prediction 
of other physical observables that also display a {\sl slowly-varying}
dynamics.  Based on this concept, we extend the GKRs to the study of
{\sl nuclear charge radii}. The GKRs are tested on 455 out of the
approximately 800 nuclei whose charge radius is experimentally
known. We find a rms deviation between the GK predictions and the
experimental values of only $0.01$~fm. This should be contrasted
against some of the most successful microscopic models that yield 
rms deviations almost three times as large. Predictions---with reliable
uncertainties---are provided for 116 nuclei whose charge radius is
presently unknown.
\end{abstract}
\pacs{21.10.Ft,21.10.Dr,21.60.$-$n}
\maketitle 

Theoretical calculations of nuclear ground-state properties may be
classified according to two primary trends. One of them is the
so-called {\sl macroscopic-microscopic} approach that is naturally
rooted in the Strutinsky energy theorem~\cite{Brack:1997}. According
to this theorem~\cite{Strutinsky:1967}, the nuclear binding energy may
be separated into two components---one large and smooth and the other
one small and fluctuating. The largest contribution varies smoothly
with both mass ($A$) and atomic ($Z$) numbers and describes the
average trend of the nuclear masses, as in the liquid drop model and
its various refinements~\cite{Myers:1969,Myers:1974}. \mbox{In contrast}, 
the small contribution fluctuates due to quantal effects ({\sl e.g.,
shell corrections}) that are often incorporated through an independent
particle model with a realistic potential that also varies smoothly
with $A$ and $Z$. The macroscopic-microscopic approach has enjoyed its
greatest success in the work of M\"oller {\it et
al.}~\cite{Moller:1993ed} and Duflo and Zuker~\cite{Duflo:1995}. The
``competing'' approach, falling under the general rubric of {\sl
mean-field} models, consists of a microscopic description in which the
average nuclear potential and the single-particle orbits are
determined self-consistently. Although mean-field models vary widely
in sophistication, their tenet is an {\sl effective interaction} or
{\sl energy density functional} that incorporates as much as possible
of the known nuclear dynamics. The effective interaction is
parametrized in terms of several empirical constants (e.g., coupling
constants and range parameters) that are then fitted to a variety of
ground-state observables.

A lesser known approach---in spite of its 40 year existence---is the
one by Garvey and Kelson~\cite{Garvey:1966zz,GARVEY:1969zz}. Rather
than attempting a global description of nuclear masses, the formalism
is based on {\sl local mass relations}. The Garvey-Kelson relations
(GKRs) have been recently revitalized both because of an interest in
understanding any inherent limitation in the nuclear-mass models as
well as due to their possible applications in stellar
nucleosynthesis~\cite{Barea:2005fz,Barea:2008zz,Morales:2009pq}. In
doing so, it was discovered that the GKRs---which are essentially
parameter free---rival in accuracy the most successful mass formulae
available in the literature~\cite{Barea:2008zz,Morales:2009pq}.

The Garvey-Kelson relations are derived from a few simple physical
principles (such as isospin symmetry) and a central assumption of a
nuclear mean field and residual interaction that vary slowly with atomic
number~\cite{Garvey:1966zz,GARVEY:1969zz}. Within this picture, Garvey
and Kelson introduced the following two local relations---each among
the masses of six neighboring nuclei---that allowed them to estimate 
an unknown nuclear mass from those of its neighbors:
\begin{subequations}
 \label{GKs}
 \begin{align}
   & \Delta M_{6}^{(1)}(N,Z)
   \equiv M(N+2,Z-2)-M(N,Z)
   \nonumber \\
   & \ \ \qquad\quad +M(N,Z-1)-M(N+1,Z-2)
   \nonumber \\
   & \ \ \qquad\quad +M(N+1,Z)-M(N+2,Z-1) = 0 \,,
  \label{GK1} \\
   &\Delta M_{6}^{(2)}(N,Z)
   \equiv M(N+2,Z)-M(N,Z-2)
   \nonumber \\
   & \ \ \qquad\quad +M(N+1,Z-2)-M(N+2,Z-1)
   \nonumber \\       & \ \ \qquad\quad +M(N,Z-1)-M(N+1,Z) = 0\,.
  \label{GK2}
 \end{align}
\end{subequations}
It is the aim of this letter to show that this method may be
successfully extended to {\sl nuclear charge radii}---a fundamental
property of atomic nuclei that is both slowly varying and has a large
experimental database~\cite{Fricke:1995,Angeli:2004}.

Valuable insights into the success of the
GKRs~\cite{Barea:2008zz,Morales:2009pq} may be gained by recalling
that according to Strutinsky's energy
theorem~\cite{Brack:1997,Strutinsky:1967}, the nuclear mass function
$M(N,Z)$ may be written as $M\!=\!\widetilde{M} + \delta M$, where
$\widetilde{M}$ and $\delta M$ denote the smooth and fluctuating parts
of the mass function, respectively. Viewing Eqs.~(\ref{GKs}) in this
light, they can be cast as $\Delta M_{6}\!=\!\Delta \widetilde{M}_{6}
+ \Delta(\delta M)_{6}$. Now, a Taylor series expansion of the
smoothly-varying (and largest) contribution $\widetilde{M}$ may be
performed around the reference point ($N,Z$) \cite{deShalit:1974} that
yields
\begin{subequations}
 \label{GKsApprox}
 \begin{align}
 \Delta \widetilde{M}_{6}^{(1)}(N,Z) &=
   \frac{\partial^{3}\widetilde{M}}{\partial Z^{2}\partial N}
  -\frac{\partial^{3}\widetilde{M}}{\partial Z\partial N^{2}}
  +\mathcal{O}(\partial^{4}\widetilde{M}) \;,
   \label{GK1Approx}\\
%
%
 \Delta \widetilde{M}_{6}^{(2)}(N,Z) &=
   \frac{\partial^{3}\widetilde{M}}{\partial Z^{2}\partial N}
  +\frac{\partial^{3}\widetilde{M}}{\partial Z\partial N^{2}}
  +\mathcal{O}(\partial^{4}\widetilde{M}) \;.
   \label{GK2Approx}
\end{align}
\end{subequations}
The above result indicates that the particular linear combinations of
masses involved in the GKRs induce strong cancellations in the Taylor
expansions---making the relations insensitive to the underlying (liquid
drop) function as well as to its first and second derivatives. Hence,
as long as successive derivatives of the underlying function become
progressively smaller (indeed, as in the liquid-drop 
formula~\cite{deShalit:1974}), the GKRs should be
satisfied to a very good approximation. Ultimately then, the level of
accuracy of the GKRs will depend on the extent to which the
fluctuating contributions $\delta M$ get cancelled out in
Eqs.~(\ref{GKs}). This we expect to be specific to the system under
consideration. In the case of atomic nuclei, the rather smooth local
behavior with $N$ and $Z$ of the mean-field potentials used to
calculate the quantum contributions---plus the fact that in
Eqs.~(\ref{GKs}) the interactions between nucleons cancel to 
first order in an independent-particle picture
\cite{Garvey:1966zz,GARVEY:1969zz}---ensures a strong 
cancellation of the quantum fluctuations among neighboring 
nuclei.

The above discussion suggests that the success of the GKRs for nuclear
masses may be extended to other observables that are driven by a
similar underlying physics. In this letter we focus on the nuclear
charge radius.  The charge radius is a nuclear structure observable
that is known with exquisite accuracy for a few nuclei in the periodic
table~\cite{Fricke:1995,Angeli:2004}. The systematic measurement of
the charge distribution of nuclei started with the pioneering work of
Hofstadter in the late 1950's~\cite{Hofstadter:1956qs} and continues
to this day with the advent of powerful continuous electron beam
facilities~\cite{Walecka:2001}. Although the experimental situation
for charge radii lags behind that of nuclear masses, an extensive
database of almost 800 charge radii already exists~\cite{Angeli:2004}.
Moreover, the advent of novel technologies and facilities to perform
electron scattering off short-lived isotopes, such as
ELISe~\cite{Simon:2007} and SCRIT~\cite{Suda:2009zz}, may extend the
data well beyond current limits in the coming years.

From the theoretical side, the most sophisticated approaches to charge
radii are based on either macroscopic-microscopic
models~\cite{Myers:1983,Duflo:1994, Buchinger:1994zz,
Buchinger:2001xg,Buchinger:2005ma} or microscopic mean-field
formulations using effective
interactions~\cite{Lalazissis:1999,Goriely:2001,Richter:2003wi, 
Samyn:2004bm,RocaMaza:2008cg,Goriely:2009mw,Goriely:2009zz}. When some
of these models are used to compare against experimental data, the rms
deviations lie in the 0.03 to 0.06~fm range~\cite{Buchinger:2005ma}.
In this letter we will show that an approach based on local relations
of the Garvey-Kelson type represents a very attractive and robust
alternative. 

Nuclear charge radii display, as in the case of masses, small
fluctuations on top of a fairly smooth average behavior
\cite{Angeli:2004,deShalit:1974}. This may be illustrated by employing
the liquid-drop inspired formula proposed in Ref.~\cite{Duflo:1994}. 
When such a formula is fitted to the charge radii of the close to 800 
nuclei included in the recent 2004 compilation by
Angeli~\cite{Angeli:2004} we obtain,
\begin{eqnarray}
   R_{ch}(N,Z) &=& 0.4980 + 0.8754\,A^{1/3} \nonumber \\ 
                    &-& 0.9845\,\alpha + 0.2703\,A^{1/3} \alpha^{2}
               ~{\rm fm} \,.
\label{LiquidDrop}
\end{eqnarray}
Here $\alpha\equiv(N-Z)/A$ is the neutron-proton asymmetry of the
nucleus and the fit produces the moderate rms deviation of 0.041~fm.
This liquid-drop inspired formula is particularly useful to estimate the
derivatives of the charge radius, as in Eqs.~(\ref{GKsApprox}).  Using
a representative set of nuclei---ranging from $^{16}$O to
$^{208}$Pb---we found the third-order derivatives of
(\ref{LiquidDrop}) to be suppressed by 4 to 6 orders of magnitude
relative to $R_{ch}(N,Z)$ itself. Moreover, given that mean-field
formulations provide quantitatively accurate predictions of
ground-state properties, we expect---as in the case of masses---strong
cancellations of the fluctuating contributions to the GKRs for nuclear
charge radii. These facts open the possibility of applying the 
GKRs to the study of nuclear charge radii.

\begin{figure}[tb]
\vspace{-0.05in}
\includegraphics[width=0.80\columnwidth,angle=0]{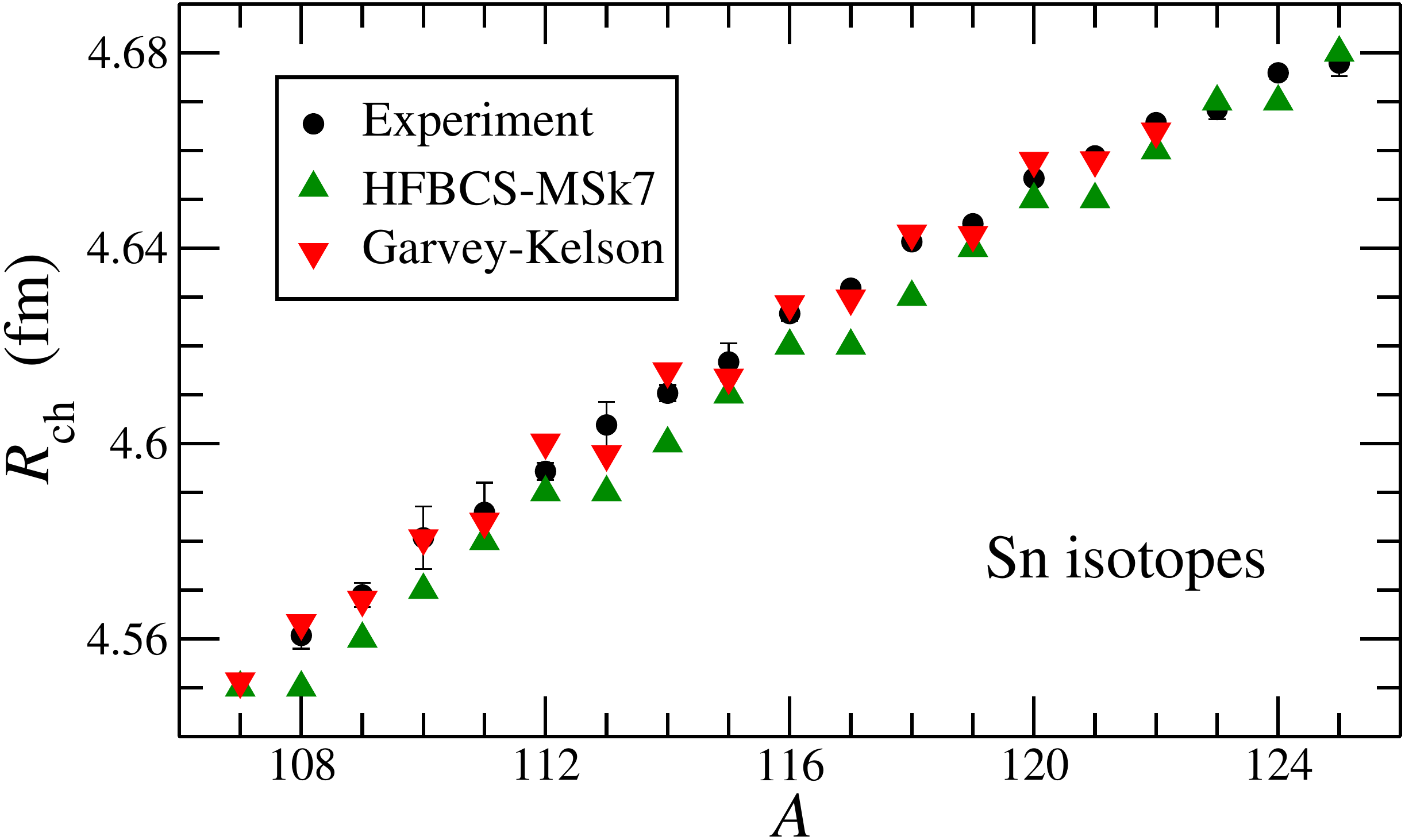}
\caption{(Color online) Comparison between theoretical
  predictions and experimental values \cite{Angeli:2004}
  for the charge radius of the Sn-isotopes.}
\label{Fig1}
\end{figure}

The implementation of the GK procedure for charge radii follows
closely the approach outlined by Barea and collaborators for the case
of nuclear masses~\cite{Barea:2008zz}. For a given nucleus, there are
(depending on the availability of experimental information) at most 12
possible estimates of its charge radius [see Eqs.~(\ref{GKs})]. All
the available estimates are then averaged to produce a GK prediction
for the charge radius of the given nucleus. The result is then
compared (when available) to the experimental
value~\cite{Angeli:2004}. In the event that the experimental value is
unavailable, a GK prediction is made for the charge radius of such a
nucleus that awaits experimental confirmation. The isotopic chain in
Tin with 18 experimentally measured charge radii provides an
illustrative example of this scheme.  Fig.~\ref{Fig1} shows the
predictions for the Tin charge radii---including the as yet unmeasured
value for ${}^{107}$Sn---using the GK relations and the
Hartree-Fock-BCS model with the MSk7 interaction~\cite{Goriely:2001}.
The HFBCS model of Ref.~\cite{Goriely:2001} generates a rms deviation
of only 0.0082~fm for these isotopes. The authors
of~\cite{Goriely:2001} have stressed that such a good agreement is
essentially parameter free, as all their model parameters were fitted
exclusively to nuclear masses. It is rewarding to see that the
GKRs---with a rms deviation of 0.0031~fm---work as good, if not
better, than the most sophisticated microscopic models available to
date. In what follows, we show that this success extends throughout
the periodic table.

\begin{figure}[tb]
\vspace{-0.05in}
\includegraphics[width=0.90\columnwidth,angle=0]{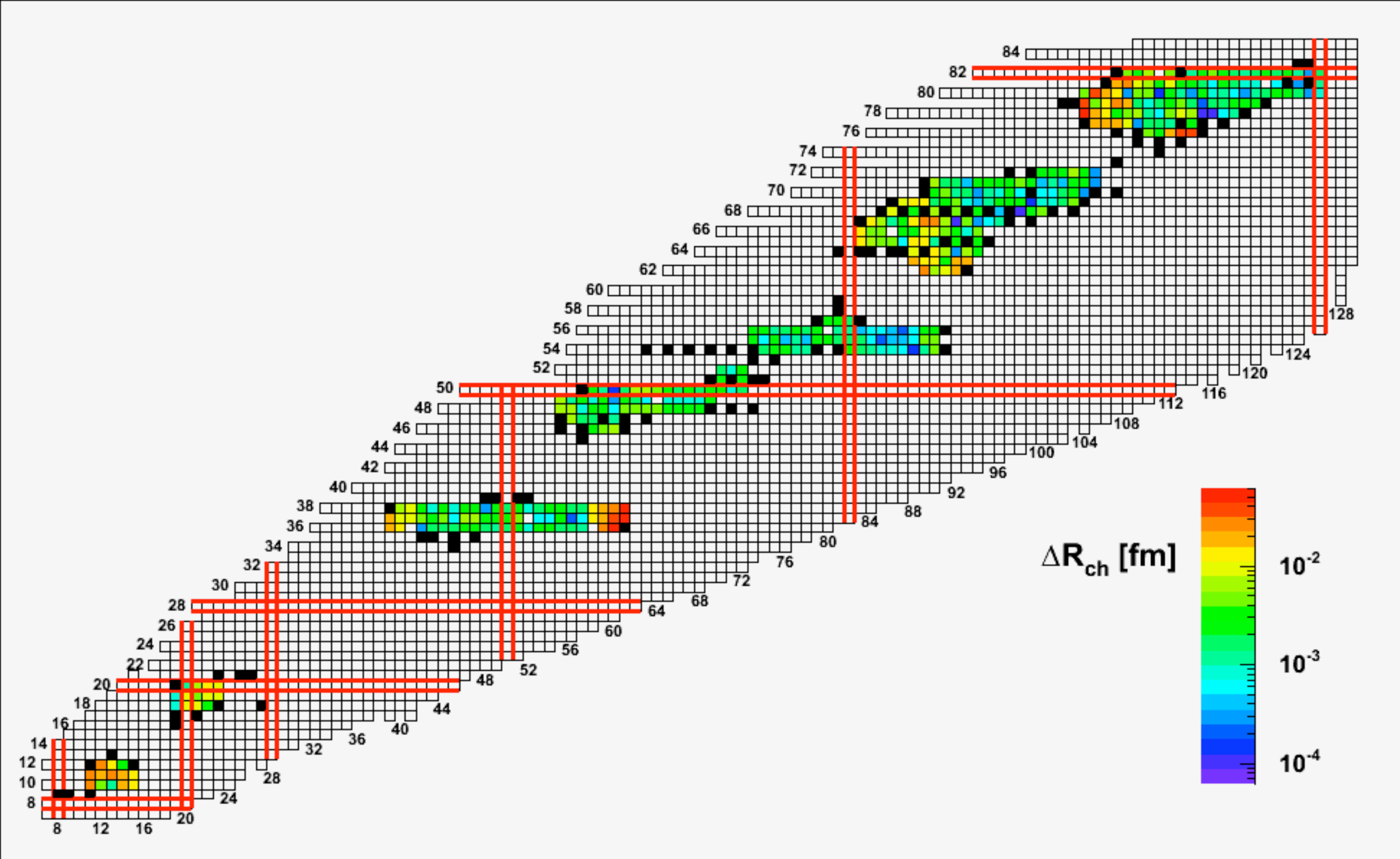}
\caption{(Color online) 
Absolute value of the difference between the GK estimate and the
experimental value for the charge radius of 455 nuclei as indicated by
the color-coded scale. Black squares denote GK predictions for 116
nuclei whose charge radius has not yet been measured. These 
predictions are provided in tabular form in Ref.~\cite{EPAPS}.}
\label{Fig2}
\end{figure}

By proceeding as in the case of the Tin isotopes, GK predictions were
made for the charge radius of a total of 571 nuclei. From these, 455
can be compared against experiment while 116 await experimental
confirmation (our GK predictions for the 116 nuclei whose charge
radius is yet unmeasured may be found in Ref.~\cite{EPAPS}). This
information has been graphically encoded in Fig.~\ref{Fig2}. The
largest deviation between the GK prediction and experiment is about
0.06~fm and this happens for only a handful of nuclei at the edges of
two of the populated regions. Most of the predictions are well below
this largest value and 331 of them fall within experimental
error. Indeed, the rms deviation obtained for the 455 charge radii is
of only $0.0097$~fm. This may be compared against the rms deviation of
$0.0275$~fm predicted by the microscopic Hartree-Fock-Bogoliubov model
HFB-8 \cite{Buchinger:2005ma} (albeit this comparison includes the 782
experimentally measured charge radii with $Z\!\ge\!8$ and
$N\!\ge\!8$~\cite{Angeli:2004}). Note that using the same experimental
data set, the new state-of-the-art HFB mass formulas
BSk17~\cite{Goriely:2009mw} and D1M~\cite{Goriely:2009zz} yield
similar rms deviations (0.030 and 0.031~fm, respectively).

\begin{figure}[tb]
\vspace{-0.05in}
\includegraphics[width=0.90\columnwidth,angle=0]{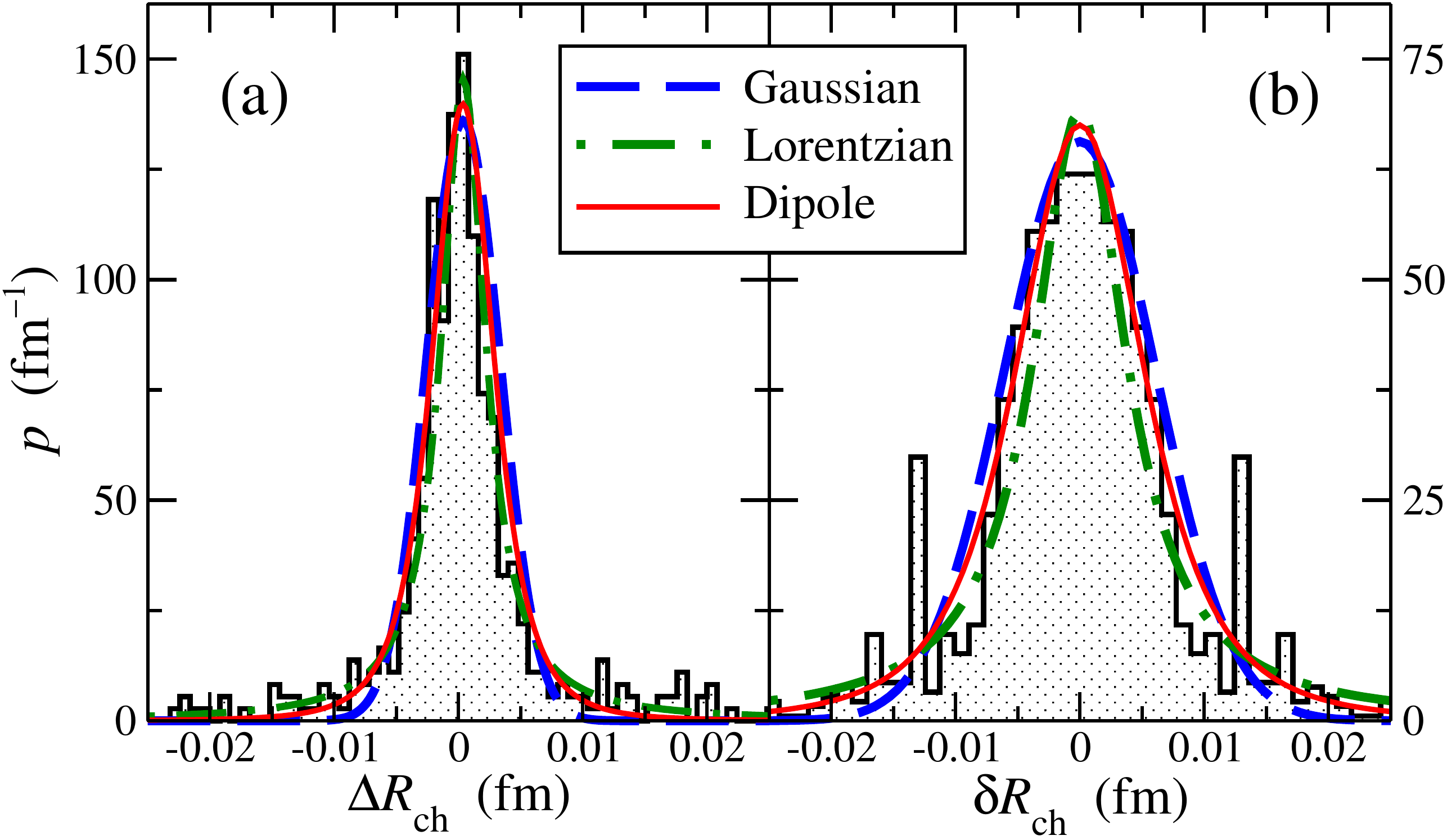}
\caption{(Color online) (a) Distribution of the differences between theory
               and experiment  ($\Delta R_{ch}$) for the charge radii of 455 
               nuclei (see text). Gaussian, Lorentzian, and Dipole
               probability density functions have been fitted to
               the histogram. 
               (b) Statistical distribution of the experimental errors
               ($\delta R_{ch}$) for the 796 nuclei included in Angeli's
               compilation~\cite{Angeli:2004}.}   
\label{Fig3}
\end{figure}

We next discuss our theoretical errors to better assess
the reliability and predictive power of the GKRs. In answering the
question of how the errors of the measured charge radii affect the GK
predictions, we avoid attaching a theoretical error by simply adding
({\sl e.g.,} in quadratures) the experimental errors associated to the
5 nuclei required to make a single GK estimate. This
method of propagating errors is uncontrolled and misleading when
applied locally, as the GKRs are satisfied with varying degrees of
accuracy throughout the nuclear chart. Rather, we adopt a global
approach that provides statistically reliable confidence levels.
The histogram in Fig.~\ref{Fig3}(a) displays the
differences $\Delta R_{ch}$ between the GK estimate and the central
experimental value for the charge radii of the 455 nuclei for which
the comparison was possible. Clearly, the probability distribution is
very narrow. To extract faithful confidence levels we have fitted
$\Delta R_{ch}$ to three different probability density functions
(PDF): the Gaussian, the Lorentzian, and the Dipole. The plot 
indicates that the Gaussian PDF falls too fast. From the remaining 
two, the Dipole gives a slightly better fit than the Lorentzian so we 
adopt it henceforth. The Dipole PDF is defined as
\begin{equation}
  p(x;\mu,\sigma)= (2\sigma^{3}/\pi) \, 
  \left[(x-\mu)^2+\sigma^{2}\right]^{-2} \;,
 \label{DipolePDF}
\end{equation}
where in the present analysis the optimal values of the mean and the
standard deviation for $\Delta R_{ch}$ are $ \mu = 3.633\times10^{-3}$
fm and $\sigma =4.554\times10^{-3}$ fm, respectively.

A particularly useful concept is that of {\sl confidence interval} of
size $n$ defined as
\begin{equation}
  {\rm CI}(n)= \int_{\mu-n\sigma}^{\mu+n\sigma} p(x;\mu,\sigma) dx\;.
 \label{ConfInt}
\end{equation}
It represents the probability that a given $\Delta R_{ch}$ will fall 
within $\pm n$ standard deviations of the mean.  For the dipole PDF, 
${\rm CI}(1)\!=\!0.818$ and ${\rm CI}(2)\!=\!0.960$. This indicates 
that the difference between the GK prediction for a nucleus 
({\sl e.g.}, any of the 116 given in \cite{EPAPS}) and the central 
experimental value falls in the range
$\Delta R_{ch}\!=\!(3.633\pm4.554)\times\!10^{-3}~{\rm fm}$
with an 82\% confidence level. We have repeated the statistical
analysis for the distribution of experimental errors. The various PDFs
are plotted in Fig.~\ref{Fig3}(b) using the same horizontal scale as
in \ref{Fig3}(a). Evidently, the experimental distribution of errors is
significantly wider. Indeed, in fitting a dipole form to it we obtain
a mean of zero (the errors are symmetric) and a standard deviation
that is more than twice as large: $\sigma_{\rm exp}\!=\!9.427
\times\!10^{-3}~{\rm fm}$. These results suggest that the GKRs provide
a useful and reliable benchmark for the calculation of nuclear charge
radii. We trust that our results may motivate the experimental
community to perform new measurements and to refine some of the
existing ones.

In summary, taking into account that the linear combinations of masses
that enter into the Garvey-Kelson relations are proportional to the
third derivatives of the slowly-varying part of the nuclear mass
function plus a remainder that comes from quantum fluctuations and is
locally small, we concluded that the GKRs could be suitably extended
to other observables obeying a similar underlying physics. In this
letter we showed that this is indeed the case for the nuclear charge radius.
Indeed, we made a systematic implementation of the GKRs using the
existing experimental database of charge radii~\cite{Angeli:2004}. Of
the 455 GK predictions that could be compared against experiment, an
overall rms deviation of only $0.0097~{\rm fm}$ was obtained.
Moreover, of these 455 predictions 331 fell within experimental error.
For comparison, one of the best microscopic models available in the
literature (the HFB-8 model of \cite{Buchinger:2005ma}) yields a rms
value of 0.0275~fm. In addition, we were able to make predictions for
116 nuclei~\cite{EPAPS} whose charge radius is presently unknown.
Finally, by performing a global statistical analysis, we attached
meaningful theoretical errors to our predictions. A similar analysis
of the experimental errors revealed a standard deviation more than
twice as large.

In the future, we want to examine to which extent the GK relations may
be of use in other finite quantum systems, such as helium and metal
clusters where the energy systematics is amenable to be described by a
semi-classical mass formula plus quantum corrections on top of
it~\cite{Brack:1997}. We also intend to extend the GK predictions to
uncharted areas of the table of nuclides with the help of methods from
the field of image reconstruction---note that important first steps in
this direction have already been taken~\cite{Frank:2007p,Barea:2007p}.
This would be particularly attractive for nuclear masses in regions of 
astrophysical interest and for charge radii of short-lived radioactive nuclei.

\begin{acknowledgments}
J.P. is indebted to Profs.\ A. Frank and J. Hirsch (UNAM) for their
hospitality and for enlightening discussions on the Garvey-Kelson
relations. J.P. also thanks Prof.\ W. Mio from the Mathematics
Department at FSU for valuable discussions on image-reconstruction
techniques. Work supported in part by grants
DE-FD05-92ER40750 (U.S. DOE), FIS2008-01661 (Spain and FEDER), and
2009SGR-1289 (Spain), and the Consolider Ingenio Programme
CSD2007-00042.  J.P. and X.R. acknowledge grants 2008PIV00094 from
AGAUR and AP2005-4751 from MEC (Spain), respectively.
\end{acknowledgments}

\bibliography{GKRadii.bbl}

\end{document}